# RPCs and readout system for the neutrino detector of the SHiP experiment

**L. Congedo**[a,b]

[a] *Università di Bari,*
  *Bari, Italy*
[b] *Sezione INFN di Bari,*
  *Bari, Italy*
  *E-mail:* liliana.congedo@ba.infn.it

ABSTRACT: SHiP (Search for Hidden Particles) is a proposed experiment to be installed at CERN, with the aim of exploring the high intensity beam frontier to investigate the so-called Hidden sector.
Since the SPS proton beam interacting with the SHiP high density target is expected to produce a large neutrino flux, the experiment will also study neutrino physics with unprecedented statistics. A dedicated Scattering and Neutrino Detector (SND) is thus being designed. It consists of a nuclear emulsion target and a tracking fibres detector in magnetic field followed by a Muon Identification System.
The Muon System is composed of iron filters interleaved with tracking planes, instrumented with Resistive Plate Chambers (RPCs) operated in avalanche mode. Each plane consists of three gaps readout by two planes of perpendicular strips.
The RPC readout electronics is being developed. It is based on the use of front-end Field Programmable Gate Arrays (FPGAs) connected to a concentration system, transmitting data serially at high speed via optical link to the data acquisition and control system.
A small-scale prototype of the SHiP Muon Identification System, with five RPC planes consisting of one large gap each, has been produced and exposed at CERN H4 in a test beam.

## Contents

## 1. The SHiP experiment

Problems unexplained by the Standard Model, such as the nature of dark matter [1], the baryonic asymmetry of the Universe [2] and the neutrino oscillations [3], could hide new physics. Several theories proposed to explain these phenomena suggest the existence of neutral fleebly interacting particles, the so-called Hidden particles.

SHiP (Search for Hidden Particles) [4] is a proposed experiment, to be installed in a beam dump facility [5] at CERN SPS (proton beam energy 400 GeV), with the aim of investigating the Hidden Sector domain (e.g. Heavy Neutral Leptons) [6]. In order to observe Hidden particles, SHiP will explore the high intensity beam frontier: $4 \times 10^{13}$ p.o.t per spill are expected (integral p.o.t. $2 \times 10^{20}$ in five years). Since the beam interacting with the fixed target (blocks of titanium-zirconium doped molybdenum alloy, density 10.22 g/cm$^3$), is expected to produce a large neutrino flux, SHiP includes a dedicated program in neutrino physics. In particular, the first direct observation of the tau antineutrino is expected, as well as the study of tau neutrino with unprecedented statistics (about $10^{16}$ $\nu_\tau$ and $\overline{\nu_\tau}$ are expected at the beam dump [7]).

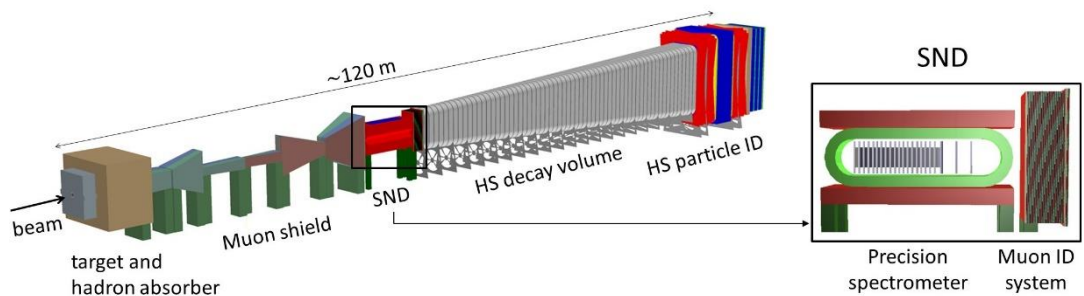

**Figure 1.** Layout of the SHiP detector (left) and of the SHiP SND (right).

The SHiP detector layout [8] (Figure 1) is composed of a hadron absorber and an active muon shield [9], deflecting magnetically muons produced by the beam interactions with the target [10], a Scattering and Neutrino Detector (SND) and a downstream Hidden Sector detector. The SND consists of a precision spectrometer, composed of a nuclear emulsion target and a tracking fibres detector in magnetic field (about 1.2 T) [11], followed by a Muon Identification System.

Downstream layers consisting of Multigap Resistive Plate Chambers (MRPCs) and acting as background tagger for the Hidden Sector spectrometer, are added to the SND.

## 2. The SND Muon Identification System

The SND Muon Identification System goal is the identification with high efficiency of the muons produced in the neutrino interactions occurring in the SND emulsion target.

The muon detector, designed by the Bari and Napoli INFN groups, consists of eight tracking planes (dimension ∼2 x 4 $m^2$), instrumented with Resistive Plate Chambers (RPCs), interleaved with iron walls, acting as hadron filters. Each tracking plane is equipped with three gaps (active area: 1.9 x 1.2 $m^2$ for each), operated in avalanche mode, since the expected rate of charged particles impinging on the detectors is about 200 $Hz/cm^2$. RPC planes are readout by two planes of perpendicular copper strips (∼1 cm pitch). Their external mechanical structure (Figure 2) includes front-end boards, high voltage and low voltage distribution, as well as the gas distribution system, studied to reduce pipes path, minimising the probability of gas leaks due to eventual pipes damages.

In order to compensate for the acceptance loss of each tracking plane due to the dead areas between adjacent gaps, the RPC planes are staggered by $\pm 10$ cm. The detector planes are hanging from the top of a Muon System support structure (Figure 2) and extracted through upper trails.

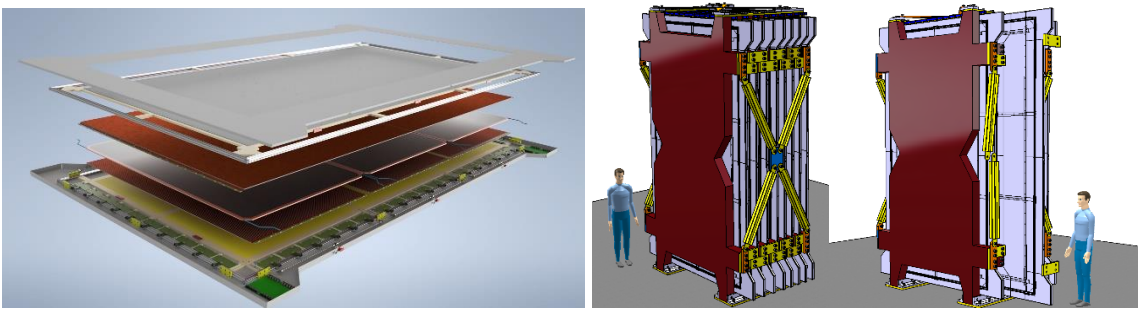

**Figure 2.** Layout of the designed Muon Identification tracking plane (left) and Muon System support structure (right).

## 3. RPC prototypes for the SHiP SND

In July 2018, the SHiP Collaboration installed a small-scale replica of the experiment target, collecting about 3 x $10^{11}$ protons at 400 GeV, in the H4 beam line at the CERN SPS, in order to measure the muon flux generated by the SHiP target [12]. The experimental setup consisted of a spectrometer instrumented with drift tubes and a muon tagger (Figure 3).

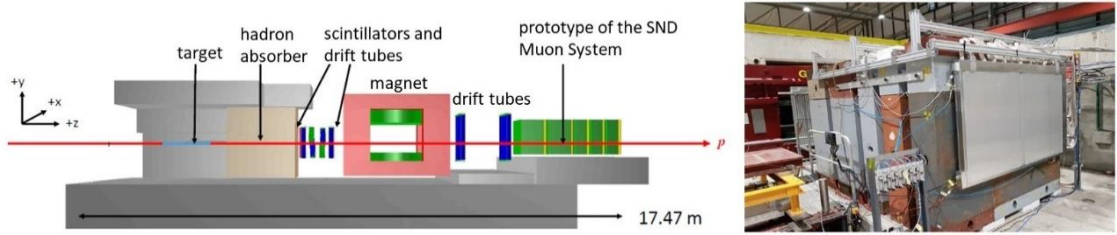

**Figure 3.** Experimental setup (left) and RPC chambers (right) used for muon-flux measurements at CERN H4 in 2018.

A pilot production of five RPCs, prototypes of the SHiP tracking planes, was made for the muon-flux measurements by Bari and Korea Universities. These detectors were interleaved with iron plates and used as muon tagger in the experimental setup.

Each RPC (active area: 1.9 x 1.2 m$^2$) consisted of 2 mm thick Bakelite electrodes covered with Graphite paint and a 2 mm wide gas gap, operated in avalanche mode with standard mixture (∼95% $C_2H_2F_4$, ∼4.5% $C_4H_{10}$, ∼0.5% $SF_6$). Chambers were readout by two panels of perpendicular copper strips with 1.0625 cm pitch and a maximum length of about 2 m.

The RPCs were both tested with cosmic rays (Figure 4) and exposed at CERN (Figure 5) by the Bari SHiP group. The efficiencies reached by the detectors were above 98% with a position resolution of about 3 mm.

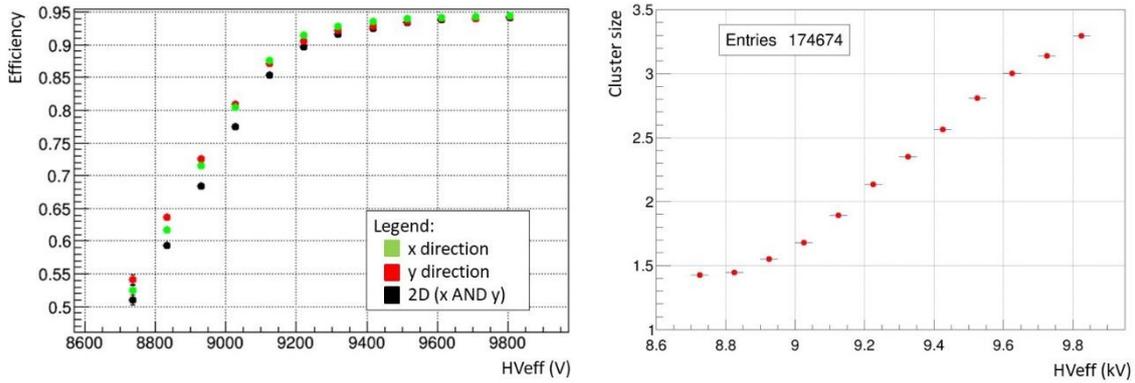

**Figure 4.** Test results of cosmic rays measurements. RPC efficiencies vs effective voltage (left). Cluster size vs effective voltage (right).

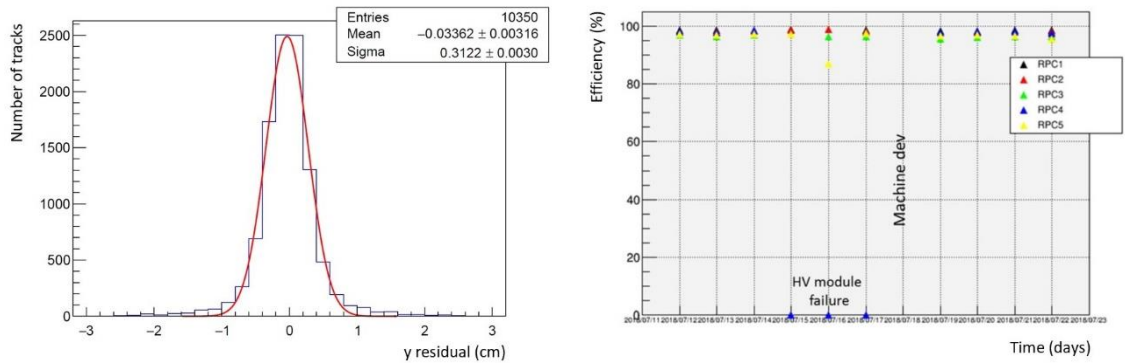

**Figure 5.** Test results of RPCs under beam exposure at CERN. RPC spatial residuals between reconstructed and effective positions (left) and detectors efficiencies vs time (right).

## 4. The RPC readout electronics for SHiP

In order to study very rare Hidden particles decays, the SHiP readout system is designed to acquire data continuously during an SPS spill (about 1 s duration), without a hardware trigger system, i.e. in triggerless mode. Data acquired by each SHiP sub-detector front-end (FE) electronics are collected by a concentration system and transmitted through optical link to a host computer in a central event filter farm for processing.

The readout electronics for SHiP RPCs (Figure 6) is being designed in Bari. It consists of a front-end system, acquiring signals in triggerless mode and transmitting data serially to the concentrators, interfaced to the SHiP data acquisition (DAQ) and control system.

The front-end electronics for SHiP RPCs is composed of 38 FE boards per tracking plane, each one connected to 16 input channels, hosting two Front-End Electronics Rapid Integrated Circuit (FEERIC) ASIC chips [13] and a FPGA. The FE ASICs acquire, amplify and discriminate signals, which are then transmitted to the FPGA onboard for data timestamping, zero suppression and serialization. The FPGA can be controlled and configured by the user through the DAQ and control system. In order to test the whole system (e.g. with cosmic rays), the trigger mode of operation is also implemented in the FPGA.

The main blocks of the designed FE FPGA are shown in Figure 6. Signals are acquired, formed according to the user configuration, zero suppressed, timestamped and stored till transmission by the Data Block. The TX Block manages transmission: it assigns packets priorities, encodes words in 8b/10b code and serializes data. The receiver (RX) Block is designed to decode and identify received data as well as to execute fast commands (e.g. trigger). Received slow control (SC) packets are managed by a dedicated block (SC Block).

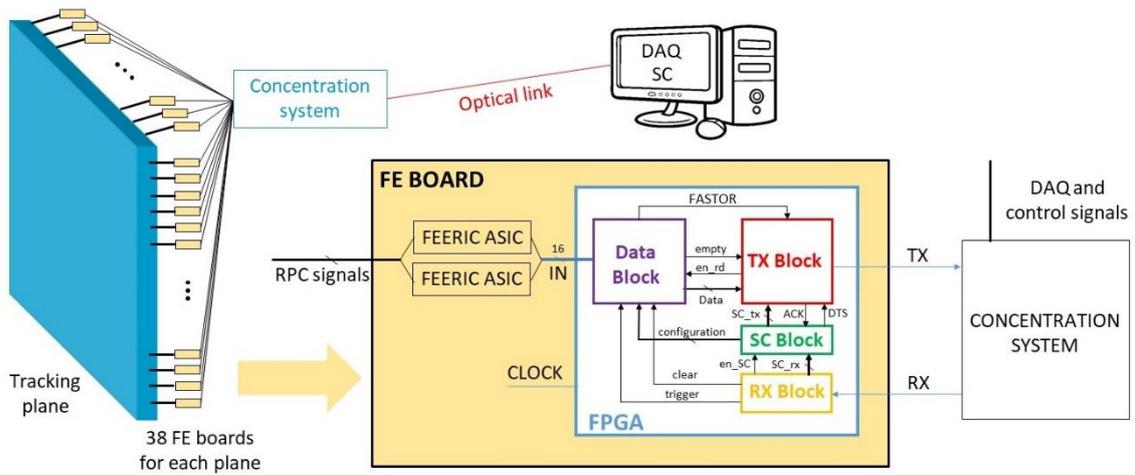

**Figure 6.** Overview of the RPC readout electronics for SHiP.

## 5. Conclusions and outlooks

A Muon Identification System, equipped with RPCs, is being designed for the SHiP Scattering and Neutrino Detector. A small-scale prototype of the Muon System has been produced, tested with standard gas mixtures and exposed at CERN H4. A position resolution of about 3 mm and efficiencies above 98% have been measured.

Eco-friendly gas mixtures for SHiP RPCs are currently under study. In order to optimize the RPCs performance with eco-gas, new gaps with reduced thickness will be produced and tested.

A front-end FPGA for the high speed serial transmission of data from the SHiP RPCs to the DAQ system has been designed and fully simulated. A prototype of the front-end board hosting the designed FPGA will be produced and tested.